\documentclass[aps,twocolumn,groupedaddress]{revtex4}
\usepackage{amsfonts}
\usepackage{amssymb}
\usepackage[cp850]{inputenc}
\usepackage{times}
\usepackage{mathptmx}
\usepackage{graphicx}
\usepackage{epsf}

\newcommand{\be}{\begin{equation}}
\newcommand{\ee}{\end{equation}}
\newcommand{\bea}{\begin{eqnarray}}
\newcommand{\eea}{\end{eqnarray}}
\newcommand{\bi}{\begin{itemize}}
\newcommand{\ei}{\end{itemize}}

\newcommand{\rkl}[1]{\left( #1 \right)}

\begin{document}
\bibliographystyle{apsrev}
\title{Application of Nonlinear Conductivity Spectroscopy to
Ion Transport in Solid Electrolytes}
\author{S. Murugavel, B. Roling} \affiliation{Institut für Physikalische
Chemie and Sonderforschungsbereich 458 (DFG), Westfälische Wilhelms-Universität Münster, Corrensstr. 30, 48149 Münster, Germany}
\date{\today}
\begin{abstract}
The field-dependent ion transport in thin samples of different glasses 
is characterised by means of nonlinear conductivity spectroscopy. AC
electric fields with strengths up to 77 kV/cm are applied to the samples, 
and the Fourier components of the current spectra are analysed. In the dc
conductivity regime and in the transition region to the dispersive
conductivity, higher harmonics in the current spectra are detected, which
provide information about higher--order conductivity coefficients. 
Our method ensures that these higher--order conductivity coefficients
are exclusively governed by field--dependent ion transport and are not
influenced by Joule heating effects. We use the low-field dc conductivity
$\sigma_{1,dc}$ and the higher--order dc conductivity coefficient
$\sigma_{3,dc}$ to calculate apparent jump distances for the mobile ions,
$a_{\rm app}$. Over a temperature range from 283 K to 353 K, we obtain
values for $a_{app}$ between 39 \AA $\;$ and 55 \AA . For all glasses, we find 
a weak decrease
of $a_{\rm app}$ with increasing temperature. Remarkably, the apparent jump
distances calculated from our data are considerably larger than typical values 
published in the literature for various ion conducting glasses. 
These values were obtained by applying dc electric fields. Our results provide
clear evidence that the equation used in the literature to calculate
the apparent jump distances does not provide an adequate physical description
of field-dependent ion transport.
\end{abstract}

\maketitle

{\bf 1. Introduction}

An important prerequisite for the application of solid ionic conductors as 
electrolytes in batteries, fuel cells and electrochromic devices is the
preparation of materials with low electrical resistance. This can be
achieved by either improving the specific ionic conductivities of materials 
or by preparing thin films. A rule of thumb for the technical applicability of
solid electrolytes is an ionic conductivity exceeding 10$^{-3}$~S/cm. 
However, in the case of thin-film materials with thicknesses of 
1 $\mu$m and less, conductivities of the order of 10$^{-6}$~S/cm to 
10$^{-5}$~S/cm can be sufficient. Thus, by means of thin-film technology, the range of
solid electrolyte materials being suitable for technical applications can be
extended considerably. In the case of crystalline materials, epitaxial
methods are often used for the preparation of thin films, while
sputtering methods and spin coating techniques are suitable for preparing
thin films of glassy and polymeric electrolytes.

When the ionic conductivity of a solid electrolyte layer does not depend on
its thickness $d$, the electrical resistance of the layer is simply proportional 
to $d$. However, in the case of thin samples, the assumption of a 
thickness-independent conductivity is often not valid. The reason for this
is the presence of large electric fields in the samples, which lead to nonlinear 
ion transport effects. Let us consider an example: A film with a thickness of $d$
= 100 nm is used in a lithium battery providing a voltage of $U$ = 5 V. In this
case, the electric field strength in the film is $E = U/d$ = 500 kV/cm. It
is well known that at field strengths $E >$ 50 kV/cm, the ionic conductivity of 
many solid electrolytes increases with increasing field strength. 
This increase of the ionic conductivity should generally improve the applicability 
of thin films. 

A possible mathematical description of the nonlinear electrical
properties of ionic conductors is the following power series for the dependence 
of the current density $j$ on the electric field strength $E$:
\be
\label{eq_power_series}
j = \sigma_1 \cdot E \,+\, \sigma_3 \cdot E^3 \,+\, \sigma_5 \cdot E^5
\,+\,...
\ee
Here, $\sigma_1$ denotes the low-field conductivity, while $\sigma_3$, $\sigma_5$ etc. 
are higher--order conductivity coefficients. Eq. (\ref{eq_power_series}) contains
exclusively odd terms, since the function $j(E)$ is an odd function.

A number of field-dependent conductivity studies have been carried out on
ion conducting glasses \cite{Vermeer56, Zagar69, Lacharme78, Hyde86,
Barton96, Isard96}. In these studies, dc electric fields,
$E_{dc}$, were applied, and the dc currents densities, $j_{dc}$, were determined. 
Often the $j_{dc}(E_{dc})$ curves can be reasonably well fitted by
a hyberbolic sine function:
\be
\label{eq_sinh}
j_{dc} \propto \sinh \rkl{\frac{q\,a_{\rm app}\,E_{dc}}{2\,k_B\,T}} \;.
\ee
Here, $q$ denotes the charge of the mobile ions, while $k_B$ and $T$
are Boltzmann's constant and the temperature, respectively. The quantity 
$a_{\rm app}$ has the unit of a distance, and is, in the following, called
an 'apparent jump distance'. Eq. (\ref{eq_sinh}) can be derived theoretically 
in the framework of a random walk theory with mobile ions carrying out 
thermally activated hops in a periodic potential landscape. In this
framework, $a_{\rm app}$ is identical to the actual
jump distance of the mobile ions, i.e. to the distance between neighboring sites. 
However, the apparent jump distances derived from fits of {\it experimental data} 
using Eq. (\ref{eq_sinh}) are generally between 15~\AA $\;$ and 30~\AA 
\cite{Vermeer56, Zagar69, Lacharme78, Hyde86, Barton96, Isard96}. 
Thus, $a_{\rm app}$ is much larger than typical distances between
neighboring ionic sites in glasses. In the framework of molecular dynamics
simulations, these typical distances were found to be about 2.5 -- 3 \AA 
\cite{Balasubramanian95, Heuer02}. 
The physical origin of the large
values for $a_{\rm app}$ is most likely related to the amorphous structure of the
matrix the ions are moving in. This amorphous structure should result in a
highly disorderd potential landscape with spatially varying site energies and
barriers, in contrast to the periodic potential landscape used in the random walk 
theory mentioned above. However, up to now, there is no comprehensive theory
relating the the shape of the potential landscape to nonlinear ion
transport.

A drawback of nonlinear conductivity studies by means of dc electric fields
is that no direct experimental information can be obtained on Joule heating
effects in the samples. Joule heating leads to an increase of the sample temperature, 
which leads again to an increase of the ionic conductivity. This
conductivity enhancement may pretend a nonlinear ion transport
effect. We would like to emphasize that the differentiation between nonlinear ion
transport and Joule heating is very important for technical applications.
While an increase of the ionic conductivity due to nonlinear transport
effects should generally improve the applicability of materials, Joule heating
often leads to an electrical breakdown of materials.

A differentiation between nonlinear ion transport and Joule heating is
possible when ac electric fields are used. According to
Eq. (\ref{eq_power_series}), the application of a sinusoidal electric field 
$E(t) = E_0 \cdot \sin (\omega\,t)$ leads to the following expression for the 
current density being in phase with the electric field, $j'$ :
\bea
\label{Eq_Fourier_components}
j' &=& \sigma'_1 \cdot E_0 \cdot \sin (\omega\,t) + \sigma'_3 \cdot (E_0)^3 \cdot 
\sin ^3 (\omega\,t) \\
\nonumber
&& + \sigma'_5 \cdot (E_0)^5 \cdot \sin ^5 (\omega\, t) + .....\\
\nonumber
&=& \sigma'_1 (\omega) \cdot E_0 \cdot \sin (\omega\,t) + \frac{3}{4} \;
\sigma'_3 (\omega) \cdot (E_0)^3 \cdot \sin (\omega\,t) \\ 
\nonumber
&& - \frac{1}{4} \; \sigma'_3 (3\omega) \cdot (E_0)^3 \cdot \sin
(3\omega\,t)\\
\nonumber
&& + \frac{10}{16} \; \sigma'_5 (\omega) \cdot (E_0)^5 \cdot \sin (\omega\,t) \\
\nonumber
&& - \frac{5}{16} \; \sigma'_5 (3\omega) \cdot (E_0)^5 \cdot \sin (3\omega\,t)
\\
\nonumber
&& + \frac{1}{16} \; \sigma'_5 (5\omega) \cdot (E_0)^5 \cdot \sin (5\omega\,t) 
+ .... 
\eea
An analoguous expression can be written down for the out-of-phase current
density $j''$ by replacing $\sigma'_i(\omega)$ by $\sigma''_i(\omega)$ and
by replacing the sine functions by cosine functions. However, in the
following, we show exclusively results for the in--phase current density
$j'$. As seen from Eq. (\ref{Eq_Fourier_components}),
the third--order term in Eq. (\ref{eq_power_series})
generates a Fourier component in the current density spectrum at
$3\,\omega$. Accordingly, the fifth-order term generates a $5\,\omega$
component, etc. Thus, the higher harmonic contributions can be used
to determine the higher-order conductivity coefficients $\sigma_3$,
$\sigma_5$ etc. In contrast, Joule heating of the sample leads to an
increase of the low-field conductivity $\sigma_1$, but does not generate
higher harmonics in the current spectrum. 

Tajitsu used this ac method to determine the higher--order conductivity
coefficients of some polymer electrolytes based on polyethlyenoxide /
lithium salt mixtures \cite{Tajitsu96, Tajitsu98}. These measurements were 
done in a frequency
range from 10$^{-2}$ Hz to 10$^6$ Hz. At low frequencies, the low-field
conductivity as well as the higher--order conductivity coefficients
become independent of frequency and identical to their dc values.
By using Eq. (\ref{eq_sinh}), Tajitsu calculated apparent jump distances $a_{\rm app}$ 
from these dc values, and he obtained values for $a_{\rm app}$
between 40 \AA $\;$ and 50 \AA . A drawback of Tajitsu's
measurements, however, is the usage of ion blocking metal electrodes. In this 
case, it is difficult to assess the respective influences on the higher-order
conductivity spectra exerted by electrochemical processes at the interfaces 
between sample and electrodes, and by electron injection into the sample.

Therefore, our approach is the utilisation of ion conducting liquid electrodes
in combination with ac electric fields and a Fourier
analysis of the current density spectra. As liquid electrodes, we use
highly conducting salt solutions containing the same type of cations,
which are mobile in the solid electrolyte. In this case, an electric field 
simply drives these cations from one liquid electrode through the solid
electrolyte into the other liquid electrode. 

By means of this method we carry out systematic temperature-- and 
com\-po\-si\-tion-dependent studies
on nonlinear ion transport effects in solid electrolytes. Very little
is known about the respective influences on nonlinear ion transport
exerted by the number density of mobile ions and by the structure of the solid 
matrix. In this paper, we present results on the nonlinear conductivity of three different 
ion conducting glasses. The compositions of these glasses are 0.127 Na$_2$O $\cdot$
0.096 CaO $\cdot$ 0.062 Al$_2$O$_3$ $\cdot$ 0.715 SiO$_2$ (NCAS12), 
0.25 Na$_2$O $\cdot$ 0.096 CaO $\cdot$ 0.062 Al$_2$O$_3$ $\cdot$ 0.592 SiO$_2$
(NCAS25), and 0.2 Na$_2$O $\cdot$ 0.8 SiO$_2$ (NS20). The composition of the first 
glass is similar to Thuringian glass 0.101 Na$_2$O $\cdot$ 0.084 CaO $\cdot$
0.03 Al$_2$O$_3$ $\cdot$ 0.785 SiO$_2$ (+ trace amounts of K$_2$O, MgO and Fe$_2$O$_3$), 
but the sodium 
oxide content is slightly higher. We chose this slightly higher sodium oxide content, since
the room temperature dc conductivity of Thuringian glass is too low to be measured 
within our temperature and frequency window. For Thuringian glass and for a 0.18 Na$_2$O 
 $\cdot$ 0.82 SiO$_2$ glass with a composition similar to our NS20 glass, 
values for the apparent jump distance $a_{\rm app}$ obtained 
from dc electric field measurements are available in the literature 
\cite{Vermeer56, Zagar69}.

By means of high--precision cutting and grinding techniques, samples with thickness between 
65 $\mu$m and 85$\mu$m were prepared. Much thinner samples are not suitable for 
our measurements, since the films have to be freestanding between the liquid electrodes.
However, since we are able to apply ac voltages with amplitudes up to 
500 V, the resulting maximum electric field amplitudes of about 77 kV/cm
are sufficient to detect higher harmonics at $3\omega$ in the current density 
spectrum. Our measurements were carried out over a frequency range 
from 10 mHz to 10 kHz and over a temperature range from 283 K to 353 K.

\vspace{1cm}
{\bf 2. Experimental}

\vspace{0.3cm}
{\bf 2.1. Preparation of thin glass samples}

Bulk glass samples were prepared from reagent grade chemicals of Na$_2$CO$_3$, 
CaCO$_3$, Al$_2$O$_3$ and SiO$_2$. All chemicals were mixed thoroughly and then 
melted at 1873 K for 5 hours. After complete homogenization, the melts were poured 
into preheated stainless steel molds with a cylindrical shape. The obtained bulk glass 
samples were then annealed 40 K below their respective glass transition
temperatures for 5 hours. 

Subsequently, the bulk samples were cut into cylindrical slices with a thickness of
about 1 mm using a high-precision cutting machine (Struers Accutom--5) equipped with a
diamond saw blade. The thickness of the slices was then further reduced by
high-precision grinding using a lapping machine (Logitec PM5). Thereby, we attained 
sample thicknesses in the range from 65 $\mu$m to 85 $\mu$m. The maximum error in the thickness
over the faces of a sample was about 2 $\mu$m. 

\begin{center}
\begin{figure}[htb]
\begin{center}
\epsfxsize=8cm\leavevmode{\epsffile{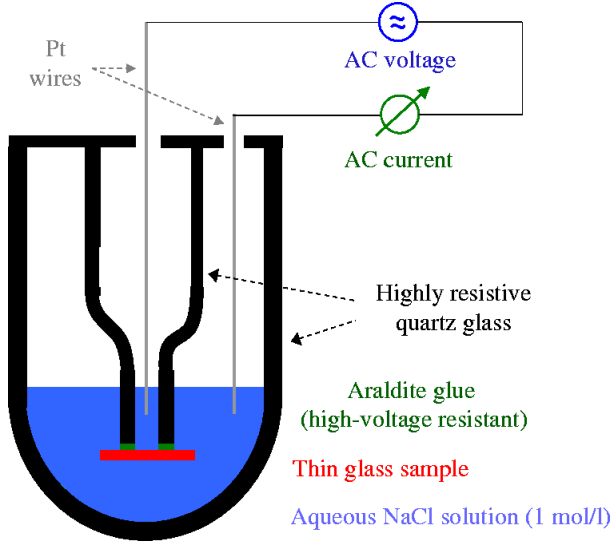}}
\caption{Schematic illustration of the 
experimental setup for nonlinear conductivity spectroscopy on 
thin glass samples}
\label{sample_setup}
\end{center}
\end{figure}
\end{center}

The thin samples were attached to a highly resistive quartz glass
tube (conductivity $< 10^{-16}$ S/cm) by means of a high-voltage resistant
Araldite glue (Vantico). 
The quartz tube with the attached sample was placed inside a quartz glass
container, see Fig. \ref{sample_setup}. 
Both the quartz glass tube and the quartz glass container were filled with a 
1 molar aqueous NaCl solution, prepared with destilled water. 
Platinum wires connected to the high-voltage measurement system
were dipped into the NaCl solutions. Since the electrical resistance
of the NaCl solution is many orders of magnitude lower than the resistance
of the glass samples, there is virtually no voltage drop in the NaCl
solutions, but the voltage applied to the platinum wires drops
completely at the glass samples. Therefore, the NaCl solutions act as
non--blocking electrodes.

\vspace{0.3cm}
{\bf 2.2. Nonlinear conductivity measurements}

\begin{center}
\begin{figure}[htb]
\begin{center}
\epsfxsize=8cm\leavevmode{\epsffile{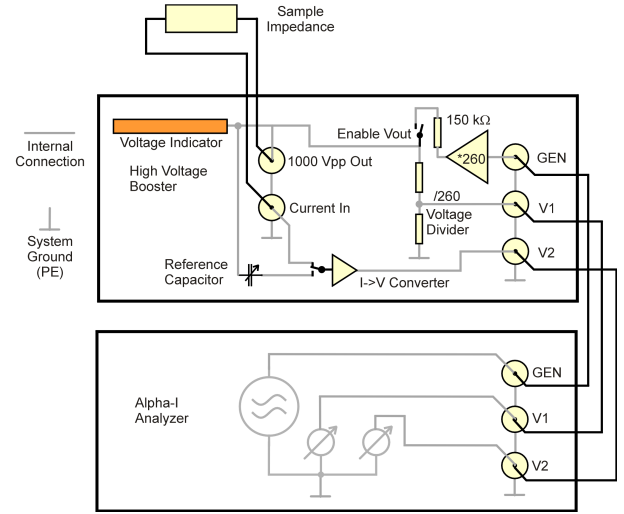}}
\caption{Schematic illustration of the Novocontrol
high-voltage measurement system.}
\label{high_voltage_setup}
\end{center}
\end{figure}
\end{center}

Our high-voltage measurement system is based on the Novocontrol
$\alpha$--S high resolution dielectric analyser, which is equipped with
a broadband high-voltage amplifier, a broadband dielectric
converter and a high-voltage sample head. A schematic illustration of
the setup is shown in Fig. \ref{high_voltage_setup}. 
The high-voltage system provides a frequency range from 3 $\mu$Hz to 10 kHz, 
a maximum voltage of 500 V, and a current resolution of 5 fA. 

The digital waveforms of sample voltage and current are numerically Fourier
transformed by a digital processor in the $\alpha$--S
analyser. Thereby, the amplitudes and phases of the base wave and 
of higher harmonics in the current spectrum are determined with respect to the 
sinusoidal voltage. Higher harmonics are detectable, if their amplitude is at least
10$^{-3}$ of the base wave amplitude.

The sample temperature was controled by the Novocontrol Quatro Cryosystem.

\vspace{0.5cm}
{\bf 3. Results}

\begin{center} 
\begin{figure}[htb]
\begin{center}
\epsfxsize=10cm\leavevmode{\epsffile{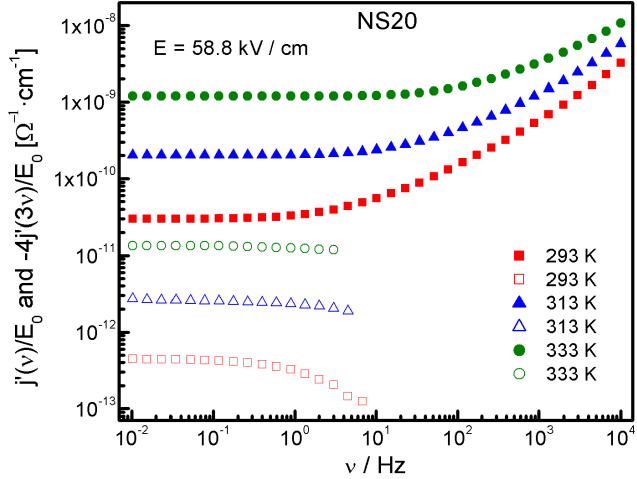}}
\caption{Base current density $j'(\nu)$ (closed symbols)
and higher harmonic  current density $j'(3\nu)$ (open symbols), 
measured after applying a sinusoidal electric field with
amplitude $E_0$ = 58.8 kV/cm to the NS20 glass. Both Fourier components of
the current density are normalised by the field amplitude $E_0$ and are plotted versus 
frequency $\nu$ at three different temperatures, $T$ = 293 K, 313 K, and 333
K. In addition, the quantity $j'(3\nu)/E_0$ is multiplied by a factor
of -4.}
\label{j_3omega}
\end{center}
\end{figure}
\end{center}

\begin{center}
\begin{figure}[htb]
\begin{center}
\epsfxsize=10cm\leavevmode{\epsffile{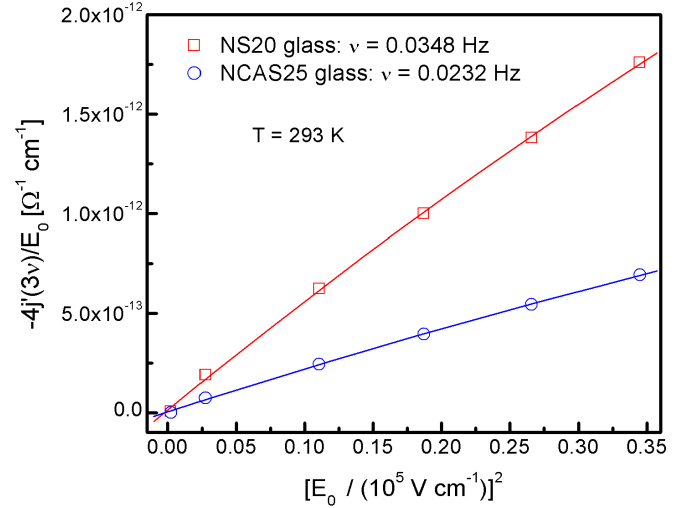}}
\caption{Plot of $-4\cdot j'(3\nu) / E_0$ versus 
$(E_0)^2$ for the glasses NS20 and NCAS25 at a temperature of $T$ = 293 K. 
The measurement errors are indicated by the size of the symbols.
The solid lines denote fits of the data by a second--order polynomial. The linear 
and the second-order term of this polynomial provide information about 
$\sigma_{3,dc}$ and $\sigma_{5,dc}$, respectively.}
\label{determ_sigma3}
\end{center}
\end{figure}
\end{center}

In Fig. \ref{j_3omega} we show results for the base current density $j'(\nu)$ 
(closed symbols) and for the higher harmonic current density $j'(3\nu)$ 
(open symbols), measured after applying 
a sinusoidal electric field with amplitude $E_0$ = 58.8 kV/cm to the NS20 glass. 
Both Fourier components of the current density spectrum are normalised by the field amplitude 
$E_0$ and are plotted versus frequency $\nu$ at three different temperatures, 
$T$ = 293 K, 313 K, and 333 K. The values obtained for $j'(3\nu)$ are
negative. According to the term $-\displaystyle\frac{1}{4} \cdot
\sigma'_3(3\omega) \cdot (E_0)^3 \cdot \sin(3\omega t)$ in Eq. (\ref{Eq_Fourier_components}), 
a negative value for $j'(3\nu)$ 
indicates a positive value for $\sigma'_3(3\nu)$. In order to include the $j'(3\nu)/E_0$ data 
in the log-log plot of Fig. \ref{j_3omega}, we have multiplied these data by a factor of -4. 
The $j'(\nu) /E_0$ curves reflect the frequency dependence 
of the low--field conductivity $\sigma'_1(\nu)$. At low frequencies, the low--field 
conductivity is independent of frequency and identical to the low--field dc conductivity.
However, at higher frequencies, there is a transition into a dispersive
regime with the conductivity increasing with increasing frequency. 
With increasing temperature, the low-field dc conductivity increases, and the transition
region from the dc conductivity to the dispersive conductivity shifts to higher
frequencies. These frequency-- and temperature--dependent features are well known for
many solid ionic conductors. 
In the dc conductivity regime and in the transition region to the dispersive
conductivity, the higher harmonic component $j(3\nu)$ was detectable, see Fig. \ref{j_3omega}. 
However, at higher frequencies in the dispersive regime, the ratio $j'(3\nu)/j'(\nu)$ 
became so small, that reliable values for $j'(3\nu)$ could not be obtained. 
The field dependence of $j'(3\nu)$ at low frequencies was further analysed 
in order to calculate the
higher--order conductivity coefficient $\sigma'_3(\nu)$. According to Eq. 
(\ref{Eq_Fourier_components}), we can write the following expression for
$j(3\nu)$:
\be
\label{Eq_j_3omega}
\frac{-4\cdot j'(3\nu)}{E_0} = \sigma'_3(3\nu) \cdot (E_0)^2 + 
\frac{5}{4}\;\sigma'_5(3\nu) \cdot (E_0)^4 + ...
\ee
Since at the applied field strengths, Fourier components at $5\omega$ in the current 
spectrum were not detectable, the second term in Eq. (\ref{Eq_j_3omega}) must be 
considerably smaller than the first one. 
Thus, when we plot $-4\cdot j'(3\nu) / E_0$ versus $(E_0)^2$, the data
should, in a good approximation, lie on a straight line with a slope of
$\sigma'_3(3\nu)$. 
As an example we show in Fig. \ref{determ_sigma3} such a plot for the NS20 and NCAS25 
glasses at a temperature of $T$ = 293 K. For this plot, the frequencies were 
chosen such that $j'(\nu)$ and $j'(3\nu)$ are identical to their dc values.
Note that in the case of the abscissa, we have normalised the amplitude of the electric 
field by the value $10^5$ V/cm. 

A closer look at Fig. \ref{determ_sigma3} reveals that the data do {\it not exactly} lie 
on a straight line, but on a curve with a slightly negative curvature. From this
we conclude that the term $\displaystyle\frac{5}{4} \cdot \sigma'_5 \cdot (E_0)^4$ in 
Eq. (\ref{Eq_j_3omega}) 
is not negligible. The solid lines denote fits of the data by a second--order polynomial. 
From the linear term in this polynomial, we obtain directly $\sigma_{3, dc}$. The quadratic 
term is negative, implying a {\it} negative value of $\sigma_{5,dc}$. Such negative values
of $\sigma_{5,dc}$ we find for all glasses and at all temperatures. However, the error in 
the quadratic terms is so large that we can exclusively determine the sign of
$\sigma_{5,dc}$, but not accurate values.

\vspace{1cm}
{\bf 4. Discussion}

\begin{center}
\begin{figure}[htb]
\begin{center}
\epsfxsize=8cm\leavevmode{\epsffile{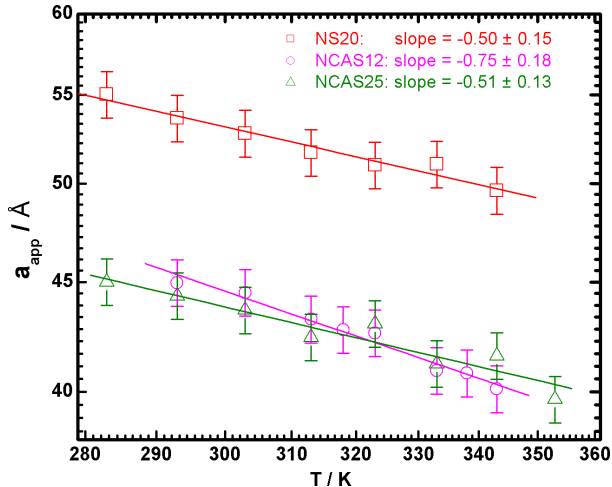}}
\caption{Plot of the apparent jump distance $a_{app}$ versus
temperature $T$ for three different glasses (log-log plot). 
The solid lines denote fits of the data to the function
$a_{\rm app} \propto T^{-\alpha}$.}
\label{a-values}
\end{center}
\end{figure}
\end{center}

Our results show that the field strengths applied in our experiments are
sufficient to obtain accurate values for $\sigma_{3,dc}$. Additionally, we
are able to determine the sign of $\sigma_{5,dc}$. In the following, we first 
compare our results to literature results obtained by applying dc electric fields. 
As mentioned in the introduction, the $j_{dc}(E_{dc})$ data of many ion
conducting glasses can, in a first approximation, be fitted by means of Eq. 
(\ref{eq_sinh}). A Taylor expansion of this equation in terms of the electric field 
yields the following expression for the ratio $\sigma_{3,dc} / \sigma_{1,dc}$:
\be
\label{eq_ratio_sig3_sig1}
\frac{\sigma_{3,dc}}{\sigma_{1,dc}} = \frac{1}{6} \rkl{\frac{q\,a_{\rm app}}{2\,k_B
T}}^2
\ee
Solving this equation for the apparent jump distance $a_{\rm app}$, we arrive
at:
\be
\label{eq_a_app}
a_{\rm app} = \sqrt{\frac{6\;\sigma_{3,dc}\;(2 k_B T)^2}{\sigma_{1,dc} \;q^2}}
\ee
This equation was used to calculate $a_{\rm app}$ for the glasses. Our measurement technique 
ensures that these values reflect exclusively nonlinear ion transport effects and 
are {\it not} influenced by Joule heating. In Fig. \ref{a-values}, $a_{\rm app}$ 
is plotted versus temperature $T$ for the three glasses. Note that both axes 
are logarithmic axes. As seen from the figure, the two glasses NCAS12 and NCAS25 
exhibit similar values of $a_{\rm app}$. At room temperature, the apparent jump
distance is about 45 \AA $\;$ and decreases to about 40 \AA $\;$ at temperatures
in the range from 343 K to 353 K.
However, in the case of the NCAS12 glass, the temperature
dependence of $a_{app}$ is slighthly stronger. The data in this log-log
plot can, in a first approximation, be fitted, by a straight line with a 
negative slope $-\alpha$. This corresponds to a power law dependence of the
apparent jump distance on temperature: $a_{app} \propto T^{-\alpha}$. For the NCAS12 glass
we find $\alpha$ = 0.75 $\pm$ 0.18 as compared to $\alpha$ = 0.51 $\pm$ 0.13 in the case of 
the NCAS25 glass. The nonlinear ion transport in the NS20 glass is characterised by
larger apparent jump distances. $a_{\rm app}$ decreases from
about 55 \AA $\;$ at room temperature to about 50 \AA $\;$ at T = 343 K. The
temperature dependence is given by $\alpha$ = 0.5 $\pm$ 0.15 and is thus
similar to that of the NCAS25 glass. 
Remarkably, the apparent jump distances $a_{\rm app}$ we have obtained from the
ratio of $\sigma_{3, dc} / \sigma_{1,dc}$ are even larger than typical values
published in the literature, which were obtained by fitting $j_{dc}(E_{dc})$ data
to Eq. (\ref{eq_sinh}). For instance, in the case of
Thuringian glass with a composition close to our NCAS12 glass, the literature 
value for the apparent jump distance is 15.5~\AA $\;$ \cite{Vermeer56}.   
Zagar and Papanikolau obtained a value
of 28.3~\AA $\;$ for a 0.18 Na$_2$O $\cdot$ 0.82 SiO$_2$ glass 
with a composition close to our NS20 glass \cite{Zagar69}. 
The reason for this apparent discrepancy between our results and
literature results is related to the validity of Eq. (\ref{eq_sinh}).
The fact that we find {\it negative} values for $\sigma_{5,dc}$ for all glasses 
shows that Eq. (\ref{eq_sinh}) does not provide an adequate physical description
of field-dependent ion transport in our glasses. 
The validity of the hyperbolic sine function would
imply {\it positive} values of {\it all} higher--order conductivity
coefficients, including $\sigma_{5,dc}$. Since this is not the case,
it is plausible that the determination of $a_{\rm app}$ from Eqs.
(\ref{eq_sinh}) and (\ref{eq_a_app}) results in different values for
$a_{\rm app}$. 

The electric field strengths applied in our study were not sufficient
to detect harmonics in the current spectra at $5\omega$. These harmonics
could be used to obtain accurate values for $\sigma_{5,dc}$. As already
mentioned, the second--order polynomial fits of the data in Fig. \ref{determ_sigma3}
yield only estimates for $\sigma_{5,dc}$. However, from these estimated values, 
we can pre-estimate that a reduction of the sample thickness by a factor 
of two should lead to detectable harmonics at $5\omega$. At the same time,
this reduction of the sample thickness should allow a further extension of
the frequency range where harmonics at $3\omega$ can be detected. Such 
a reduction of the thickness should be achievable by means of chemical
etching techniques. 

From a theoretical point of view, it is a great challenge to obtain a 
better understanding of 
the large apparent jump distances $a_{app}$,  of their temperature dependence, 
and of the negative sign of $\sigma_{5,dc}$. We would like to emphasize 
that an apparent jump distance of the order of 50 \AA $\;$ does {\it not} imply 
that ions perform single jumps over such a large distance. In the dc conductivity
regime, we do not probe local hops between neighboring sites, but we probe
the long-range ion transport. In the case of a disordered potential
landscape with site energy disorder, the distances relevant for the long--range 
ion transport could be distances between low--energy sites in the landscape
and/or distances between sites close to the percolation energy. 
Detailed theoretical studies on particles moving in disordered potential
landscapes will be required in order to obtain a detailed microscopic understanding
of these relevant distances and their influence on the apparent jump
distance $a_{\rm app}$. Such studies are currently underway \cite{Heuer04}.

\vspace{1cm}
{\bf 5. Conclusions}

We have measured the nonlinear ionic conductivity of three different
glasses in a temperature range from 283~K to 353~K. In the dc conductivity
regime and in the transition region to the dispersive conductivity, the
application of ac electric fields with amplitudes up to 77 kV/cm leads to
harmonics in the current spectrum at $3\omega$, which can be used to calculate
the higher--order conductivity coefficient $\sigma'_3$. From the ratio $\sigma_{3,dc} 
/ \sigma_{1,dc}$, we derive values for the apparent jump distance $a_{\rm app}$ of the
mobile ions. In the temperature range of our study, we find values
between 39 \AA $\;$ and 55 \AA. The temperature dependence
of the apparent jump distance is, in a first approximation, given by
$a_{\rm app} \propto T^{-\alpha}$, with $\alpha$ values ranging
from 0.5 to 0.75. Our values for $a_{\rm app}$ are significantly higher
than literature values for various glasses obtained by fitting 
$j_{dc} (E_{dc})$ data to the hyperbolic sine function 
$j_{dc} \propto \sinh (q\,a_{\rm app}\, E_{dc} / (2\,k_B T))$.
One reason for this apparent discrepancy is the fact that the hyperbolic sine 
function does not provide an adequate physical description of field-dependent
ion transport. This can directly be concluded from the {\it negative}
sign of $\sigma_{5,dc}$, which we find for all glasses. We pre-estimate that a further 
reduction of the sample thickness by a factor of two, which should be achievable 
by means of chemical etching techniques, will allow us to considerably extent the
frequency range, where harmonics at $3\omega$ can be detected, and, in
addition, to obtain accurate values for $\sigma_{5,dc}$. 

\vspace{1cm}
{\bf Acknowledgements}

We are indebted to A. Heuer for many stimulating discussions,
and to S. Pas for critically reading the manuscript.
Furthermore, we would like to thank the Deutsche Forschungsgemeinschaft 
(SFB 458) and the Fonds der Chemischen Industrie for financial support 
of this work.

\end{document}